\documentclass[9pt,twocolumn,twoside]{osajnl}
\journal{jocn}
\usepackage{amsmath,amsfonts}
\usepackage{algorithm}
\usepackage{array}
\usepackage{textcomp}
\usepackage{stfloats}
\usepackage{url}
\usepackage{verbatim}
\usepackage{graphicx}
\usepackage{color, soul}
\usepackage{graphicx} 
\usepackage{booktabs}
\usepackage{todonotes}
\usepackage{caption}
\usepackage[caption=false,font=normalsize]{subfig}

\setboolean{shortarticle}{false}

\title{Sparse Distributed Fiber Optic Sensor Placement in Metropolitan and Datacenter Elastic Optical Networks}

\author[1,*]{Sleman Mouammar}
\author[1]{\'Italo B. Brasileiro}
\author[2]{Andr\'e Drummond}

\affil[1]{Technische Universit\"at Braunschweig, Germany}
\affil[2]{Department of Computer Science, University of Bras\'ilia, Brazil}

\affil[*]{sleman.mouammar1@tu-braunschweig.de}

\begin{abstract}
Distributed Fiber Optical Sensing (DFOS) leverages optical backscattering signals to predict failure events and enable proactive Just-In-Time (JIT) restoration. Modern network infrastructures have optical components and fibers as key enablers of high throughput and low latency. The integration of sensing features in this optical infrastructure adds features such as network diagnosis, spatial positioning and sensing, and opens up opportunities for the implementation of intelligent network control expected for the next-generation networks. In this paper, we study how the quantity and the placement of the DFOS devices in an all-optical network can mitigate failure events. In our evaluation, we consider metropolitan and data center elastic optical networks. This work also evaluates sensing devices with various prediction time capabilities, in order to assess their effectiveness when considering a variable number of deployed devices and different placement heuristics. The evaluation uses key performance metrics, including the bandwidth blocking rate, number of affected circuits by failure events, and the service downtime. Simulation results demonstrate that the Greedy Shortest Path Coverage (GSPC) heuristic substantially reduces the number of circuits affected by failure when DFOS devices cover only 5\% of links, while 25\% coverage shows a performance gain comparable to the full coverage scenario, which is an opportunity for significant cost reduction. Finally, for the 5\% sensing coverage scenarios, our results illustrate that prediction windows of 15~ms and 7.5~ms are sufficient for JIT restoration in metropolitan and data center networks, respectively.

\end{abstract}

\setboolean{displaycopyright}{false} 

\begin{document}

\maketitle

\section{Introduction}

Integrated Sensing and Communication (ISAC) represents an emerging paradigm that unifies environmental sensing and data communication in a shared waveform, spectrum, and hardware infrastructure. ISAC enables capabilities such as object detection, tracking, recognition, behavior analysis, and environmental awareness. ISAC is gaining traction in industrial applications among telecommunications vendors, including Huawei in vehicular networks, Ericsson in smart urban infrastructure, and Qualcomm in autonomous driving and IoT systems \cite{zhang2026integrated}. Furthermore, standardization bodies such as 3GPP, ETSI, and IEEE \cite{itu2025g681} are formalizing concepts to support the expanding range of ISAC use cases anticipated for 6G networks, in which network infrastructure sensing and monitoring is a key requirement when providing services such as fast recovery and high-availability.

\begin{figure}[h!]
        \centering
        \includegraphics[width=1\linewidth]{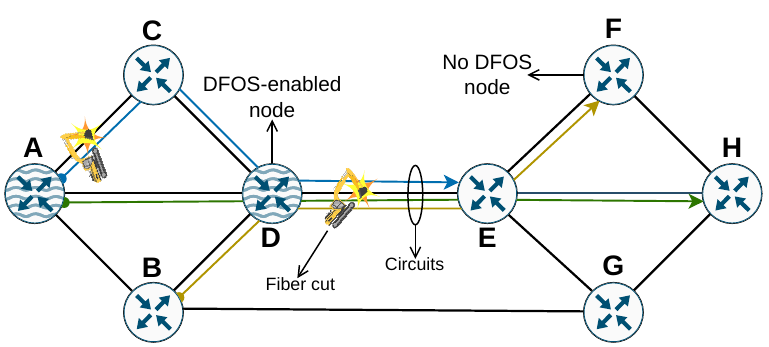}
        \caption{DFOS sparse distribution in an 8-node network: scenario with sensors placed at nodes \emph{A} and \emph{D}, and failure events (fiber cuts) to take place at unidirectional links $A\rightarrow C$ and $D\rightarrow E$.}
        \label{fig:DFOS placement}
    \end{figure}

Distributed Fiber Optic Sensing (DFOS) enables the integration of ISAC into the network by providing sensing capabilities for acoustic, vibration, temperature, and strain, and thus enabling communication with additional sensing services, such as traffic and security monitoring, fire monitoring and prevention, detection of water leakage, and structure change, as reported in \cite{wellbrock2023explore}. DFOS systems read back-scattered optical signals in fibers to function as continuous sensors that monitor tens of kilometers of infrastructure. Because of that, the deployment of high-precision DFOS technology has the potential to replace thousands of local discrete point sensors across network links, bringing significant Capital Expenditure (CapEx) savings. In order to further optimize cost savings, DFOS devices can be strategically positioned at sparse locations, with the aim of maintaining a good monitoring coverage with reduced number of deployed sensing devices. This sparse DFOS placement problem has been formally proven to be NP-hard \cite{ye2022distributed}, and it is affected by multiple factors, such as the traffic pattern, the topological layout, and the sensing device efficiency. While sensor placement has been addressed in the literature, the impacts of such deployment strategies on network performance and survivability are still a research gap. Figure \ref{fig:DFOS placement} exemplifies the sparse DFOS placement problem.

Fig.~\ref{fig:DFOS placement} illustrates a network with two DFOS-enabled nodes deployed to monitor failure occurrence in a set of unidirectional links ($A\rightarrow C$ and $D\rightarrow E$), further enabling circuit restoration. In the depicted snapshot, one circuit $A\rightarrow C\rightarrow D\rightarrow E$ (blue) traverses the link $A\rightarrow C$, and three circuits, $A\rightarrow C\rightarrow D\rightarrow E$ (blue), $A\rightarrow D\rightarrow E \rightarrow H$ (green) and $B\rightarrow D\rightarrow E \rightarrow F$ (yellow) traverse the bottleneck link $D\rightarrow E$. In both links, a failure is about to occur. If the DFOS device is deployed only at node $A$, the proactive restoration will only reallocate one-third (blue) of the affected circuits, while the remaining circuits will experience disruption. In the case of site $D$ holding a DFOS, the failure on the central link is detected in advance, and all three affected circuits are proactively reallocated, reducing the impact of service disruption. In previous work~\cite{mouammar:26}, we exploited DFOS as an enabler for Just-In-Time (JIT) (proactive) restoration to demonstrate that one could achieve high network resilience to link failures without paying the resource over-allocation costs of a protection mechanism. In that study, DFOS were enabled in all network nodes, which is a strong assumption. The results showed a circuit disruption reduction by more than 90\% in a metropolitan Elastic Optical Network (EON).

This paper moves further to investigate the impact of sparse DFOS placement on failure prediction, which is a practical deployment scenario that will be considered in real networks. The main contribution is threefold: (i) we evaluate the impact of DFOS sparse placement for failure prediction in metropolitan and data center (DC) EONs, (ii) we adapt a set of heuristics for DFOS sparse placement and propose the Greedy Shortest Path Coverage (GSPC) solution as an attempt to mitigate the impact of failures, and (iii) we analyze the impact of deploying different types of DFOS based on different failure prediction window capability. 
Our findings indicate that, besides reducing the impact of failures, different DFOS sparse placement solutions significantly influence the outcomes. Simulation results further show that deploying DFOS devices to monitor only 25\% of network links has performance comparable to a network fully covered (sensed) by DFOS, which results in significant CapEx savings. Furthermore, DFOS devices with a 15~ms failure prediction window provide the most effective and efficient service maintenance, even with only $5\%$ of monitored links.

The rest of the paper is organized as follows: section \ref{sec:literature} presents the related work, section \ref{sec:arch} details the proposed architecture, section \ref{sec:heuristics} presents the DFOS placement heuristics, section \ref{sec:perfEval} discusses the numerical results, and section \ref{sec:conc} presents conclusions and related future work.


\section{Related work}\label{sec:literature}

Network operators aim to reduce both hardware and operational costs by enhancing network availability and minimizing service disruptions. DFOS technologies can contribute to these goals, however, their effectiveness depends on the adoption of appropriate sensor placement strategies. In \cite{ye2022distributed}, a DFOS placement method is proposed, in which a single sensor can monitor multiple fiber links, with a total sensing length of up to 80 km, resulting in a 26\% improvement compared to a random-fit baseline. In addition, \cite{ye2021survivable} examines DFOS placement under single-link failure scenarios in the context of protection mechanisms, using the same sensing-length assumption. The reported results show a 16\% reduction achieved by the best solution (explore-and-pick) in the number of sensors deployed to ensure full network coverage. In this paper, we study alternative DFOS placement models and compare several heuristic strategies for sensor placement. DFOS devices will operate proactively to mitigate the impact of network failures by means of JIT restoration. We consider a simplified sensing model in which each DFOS is placed at a source node and monitors a single unidirectional outgoing link.

Network survivability studies often leverage Machine Learning (ML) to predict failure events. Optical performance monitoring at the transmitter associated with State of Polarization (SoP) data along the lightpath provides detailed information on signal quality degradation. These measurements can serve as input features for ML models, supporting proactive failure prediction. In \cite{10901306}, a Semantic FaultAware (SFA) architecture is proposed to enable proactive fault prediction in Wide Area Networks (WANs) by adding a semantic layer above the control plane. Experiments on real network data demonstrate that ML techniques can predict potential faults with up to $88\%$ accuracy. Additionally, the work in \cite{11435207} performs a field trial of multiple modalities (backscattering, forward transmission, and hybrid) of distributed sensing devices in a metropolitan network, with a proposed AI framework to detect vehicle activity, construction induced vibration, and even weak events such as footsteps. Finally, the work in \cite{ambrosone2026enabling} proposes a telemetry framework that leverages high-sampling sensing data and Apache Kafka to decouple the sensing data stream from storage to immediate AI/ML agent consumption, enabling correlation of heterogeneous sensing metrics in real-time. Considering the high efficiency of ML solutions in predicting failures, the experiments in this paper assume the deployment of such trained models, in order to enable the evaluation of sensing event detection on network performance. Furthermore, any AI/ML-based approach could be applied to enhance the results obtained here.

Network protection can significantly benefit from optical switching devices with high reconfiguration speeds. While the ITU-T standard specifies a maximum restoration time of 50 ms for protection mechanisms, Wavelength Selective Switches (WSS) are capable of much faster switching, typically in the range of 10–20 ms. The impact of leveraging such devices for failure recovery is investigated in \cite{he202550}, which proposes a shared protection heuristic for large-scale Wavelength-Division Multiplexing (WDM) networks with time-constrained resource allocation to ensure path recovery within 50 ms. In contrast, our work examines the effect of devices with varying reconfiguration times and analyzes how variations in both their quantity and placement mitigate the impact of failures.

As presented in \cite{11424028}, ISAC can be established in different mediums (e.g. fiber, free space, visible light) by using different principles, such as forward and/or backward sensing mechanism. One key challenge in providing efficient ISAC is to design an optimized system and sensing architecture, enabling high spectral efficiency. The work in \cite{11424028} surveys sensing architectures and recent advances in different systems such as WDM, Frequency-Division Multiplexing (FDM), Space-Division Multiplexing (SDM) and Time–Frequency Multiplexing (TFM). In this paper, we built on top of the ISAC systems defined in \cite{11424028} and \cite{itu2025g681} to propose an adapted sensing architecture to EON, which is later experimented in metropolitan and data center networking.


The work in \cite{szczerban2024optical} introduces a Fiber Sensing Control Device (FSCD) featuring a control plane that enables the tight integration of sensing and communication. The study addresses the strategic placement of FSCDs, enabling the creation of sensing regions that are covered by joint SoP- and Optical Time Domain Reflectometer (OTDR)-based monitoring. The proposed system achieves local decision-making latencies below 50 ns and sensing responsiveness under $1~\mu$s, showing the rapid feedback of such sensing devices. This paper assumes the deployment of such sensing devices and examines the impact of their optimal placement on failure mitigation.


\section{DFOS-enabled architecture in EON}\label{sec:arch}


\begin{figure*}[ht]
    \centering
    \includegraphics[width=1\linewidth]{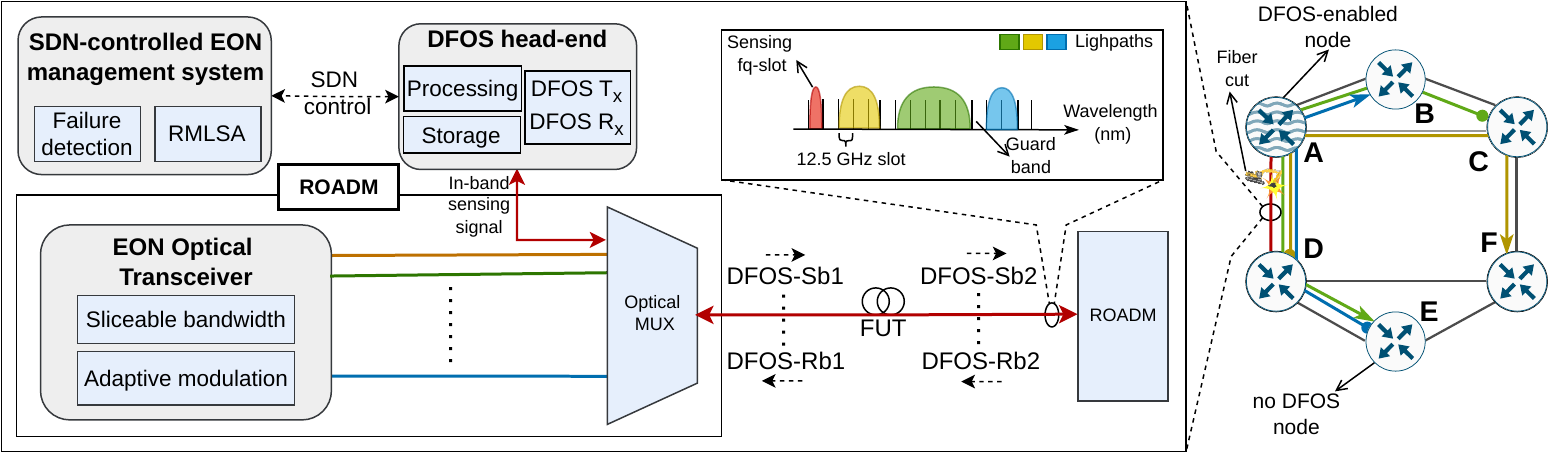}
    \caption{DFOS-enabled system architecture  for EON based on back scattering sensing (BSS) and in-band sensing frequency-slot, where the right side illustrates an exemplary network with one DFOS-enabled node, where all lightpaths traversing the sensing-enabled link ($A\rightarrow D$) are sensed. On the left hand side, is a zoom-in to the DFOS architecture.}
    \label{fig:Sensing_device}
\end{figure*}

Built on top of the Dense-WDM (DWDM) sensing architecture specified in ITU-T standard~\cite{itu2025g681}, Figure \ref{fig:Sensing_device} presents an adapted architecture for a 6-node elastic optical network, where a DFOS device is placed at source node $A$, thus allowing the monitoring of a single unidirectional link (red link $A \rightarrow D$) and the sensing of all the lightpaths (i.e., green, yellow, and blue) traversing that link. The deployed DFOS enables Back-Scattered Sensing (BSS) using an in-band sensing frequency slot in the spectral grid. The proposed architecture is designed for metropolitan and data center optical networks and enables distributed monitoring of the physical fiber infrastructure, while maintaining simultaneous data transmission over the same fiber. The EON sensing architecture provides real-time and distributed awareness of fiber conditions through a DFOS subsystem integrated with EON optical transceivers, elastic optical components, flex-grid reconfigurable optical add-drop multiplexers (ROADMs), and a software defined network (SDN)-enabled control plane. The architecture supports continuous fiber monitoring and failure-aware network survivability through JIT restoration. In contrast to an out-of-band sensing architecture, the proposed approach allocates a dedicated sensing frequency slot within the EON spectrum, allowing the sensing signal to propagate simultaneously with data lightpaths over the fiber under test (FUT).

The DFOS subsystem consists of a DFOS head-end, an in-band sensing signal, and the optical coupling and control components. The DFOS head-end comprises a sensing transmitter (DFOS Tx), a sensing receiver (DFOS Rx), and processing and storage modules. The DFOS Tx generates short optical sensing pulses that are launched into the FUT through the reserved sensing frequency slot. The sensing signal propagates along the monitored fiber and generates Rayleigh backscattered signals that propagates back to the source. The backscattered signal is received by the DFOS Rx and processed to obtain spatially resolved information about the physical condition of the FUT. The processing module extracts disturbance characteristics, such as vibrations, acoustic events, bending, and fiber-cut events, while the storage module maintains the resulting measurements and timestamps for subsequent correlation with network events.

The sensing signal is coupled into the EON optical fiber through an optical multiplexer (MUX), as illustrated in Fig.~\ref{fig:Sensing_device}. The optical MUX combines the DFOS sensing signal with the data channels generated by the EON optical transceiver. The resulting signal is the combination of multiple spectrally separated channels, which is transmitted through the FUT toward the next ROADM. The ROADM provides flexible wavelength-selective switching of the optical spectrum, but the dedicated sensing frequency slot maintains fixed bandwidth and central frequency. This architecture therefore enables the DFOS signal and elastic data lightpaths to share the same physical fiber without requiring an out-of-band sensing fiber. 

By exploiting the flexible-grid capability of wavelength-selective switches, frequency slots with a granularity of 12.5~GHz are sufficient to hold narrow-band distributed acoustic sensing probe signals, which typically operate at the MHz level. Consequently, the sensing signal can coexist with elastic data channels while introducing only minimal spectral overhead~\cite{wang2026sensing}. Accordingly, the DFOS sensing signal occupies a singular reserved in-band frequency slot of 12.5~GHz, fixed as the first index within the elastic optical spectrum. As illustrated in Fig.~\ref{fig:Sensing_device}, the sensing slot is spectrally separated from neighboring lightpaths by a guard band to limit interference between sensing and data signals. This enables simultaneous data transmission and distributed fiber sensing over the same optical infrastructure. 

Four logical optical reference points are defined along the FUT: DFOS-Sb1 and DFOS-Rb1 denote the sensing transmit and receive reference points at the DFOS-side interface of the monitored fiber, while DFOS-Sb2 and DFOS-Rb2 are the corresponding reference points at the remote network-side interface. These reference points define the boundaries of the sensing domain and provide a consistent basis for specifying the optical characteristics of the sensed lightpaths inside the link ($A \rightarrow D$) independently of the underlying network topology.

Building on this DFOS-enabled architecture, we consider a failure-aware, SDN-controlled EON management system for metropolitan and data center networks that supports accurate failure event detection, localization, and survivability through JIT restoration. As depicted in the top-left of Fig.~\ref{fig:Sensing_device}, the management system comprises two functional modules: a failure detection module, which consumes the disturbance features reported by the DFOS head-end, and a routing, modulation level, and spectrum allocation (RMLSA) module, which computes path and spectral resources
for the affected lightpaths. Both blocks act on the optical layer through an SDN control interface towards the ROADM and the DFOS head-end. 

To illustrate how the proposed system operates, upon detecting a failure, the DFOS head-end processes the backscattered signal to determine the location of the event along the FUT. The resulting failure information is forwarded to the SDN controller, which identifies the affected lightpaths and utilizes JIT restoration to initiate the reconfiguration of the affected optical connections as soon as the failure information becomes available, thereby reducing the service disruption time. Using the example in Fig.~\ref{fig:Sensing_device}, JIT restoration is enabled in fiber $A\rightarrow D$ by the monitoring DFOS device deployed in node A. In this case, a failure in link $A\rightarrow D$ can be suppressed with JIT restoration for a seamless recovery, but a failure on link $D-E$ would cause service interruption and delayed recovery, as no DFOS is deployed in any of the nodes $D$ or $E$.
The main goal of this work is to study the impact of deploying DFOS devices and how their placement affects service disruption and recovery in single-failure scenarios. This work considers that failure prediction is 100\% accurate, thus, false positives are not within the scope of this study.

In order to assess the impact of circuit reallocation signaling following a failure, the overall circuit downtime performance metric is utilized. As defined in Eq.~(\ref{eq: downtime}), the downtime consists of two main components: the cumulative setup time ($ST$) over all allocation attempts ($a$), and the switching time ($SWT$)~\cite{765615},

\begin{equation}
\text{Downtime} = a \cdot ST + SWT
\label{eq: downtime}
\end{equation}

In restoration, the SDN-enabled control plane searches for an alternative path after detecting the failure event, where every allocation attempt (a) incurs a setup time $ST$, defined as

\begin{equation}
\text{$ST$}=  n \cdot P + (n + 1) \cdot D ,
\label{eq:setuptime}
\end{equation}

in which $n$ is the number of hops in the primary path, $P$ is the propagation delay, and $D$ is the node processing delay. Due to the reactive nature of restoration, the switching time $SWT$ accounts only for signaling along the broken primary path up to the failed link and is given by Eq. \ref{eq:pst_restoration}:

\begin{equation}
\text{$SWT$} = F + n \cdot P + (n + 1) \cdot D,
\label{eq:pst_restoration}
\end{equation}

where $F$ is the failure detection time.

\section{DFOS Device Placement Heuristics}\label{sec:heuristics}
Since the sparse DFOS placement problem is NP-hard \cite{ye2022distributed}, we now present a heuristic based on previously proposed solutions for the regenerator distribution problem in optical networks \cite{chaves2012novel}. In order to verify the efficiency of each specific placement solution, we vary the number of DFOS devices deployed in two EON scenarios: data-center and metropolitan optical networks. The set of evaluated heuristics includes: (i) Most Used (MU) links, (ii) Most Simultaneously Used (MSU) links, and (iii) Node Degree First (NDF), inherited from the regenerator placement problem; (iv) Random links approach to serve as baseline, and (v) the proposed heuristic GSPC.

MU and MSU heuristics distribute the DFOS devices according to the most used links. For both solutions, a preliminary simulation is required to evaluate the traffic profile and provide a link list ordered by utilization, which is used by both heuristics to balance the DFOS distribution. The main difference between MU and MSU is in the rule applied to build the utilization list derived from the preliminary simulation. The MU heuristic uses a counter for each link in the network, which is incremented whenever a new established circuit crosses that link. Thus, at the end of the preliminary simulation, the link list will measure the total number of circuits that crossed each link in the network. This list is ordered in descending order, and the top $N$ links will have DFOS deployed, according to the total number $N$ of DFOS available. On the other hand, MSU builds a similar list, but now it will store the highest number of simultaneously active circuits in each link, different from the total stored by MU.

The NDF algorithm does not require a preliminary simulation, and the distribution is based solely on the node degree of all the nodes in the network. The NDF sorts the list of nodes based on the highest degrees in descending order and then chooses the links from top to bottom according to the number $N$ of available DFOS devices. The Random Links heuristic, as the name suggests, randomly selects a set of $N$ links to monitor using the $N$ DFOS devices available. More details and the pseudocode for MU, MSU, and NDF solutions can be found in \cite{chaves2012novel}.

We now introduce the GSPC algorithm; its main idea is to achieve high-efficiency placement of DFOS devices and enable faster and more effective JIT restoration. By exploiting the deployed K-shortest path (KSP) routing solution, the GSPC heuristic prioritizes links that participate in multiple shortest paths, thus increasing coverage of the sensing capability and the chances of reallocating more circuits. The Algorithm \ref{alg:sp_coverage} describes the selection of links to be sensed with DFOS devices.

\begin{figure*}[hb]
    \centering
        \subfloat[ION topology \cite{ion}.]{
        \includegraphics[width=0.33\textwidth]{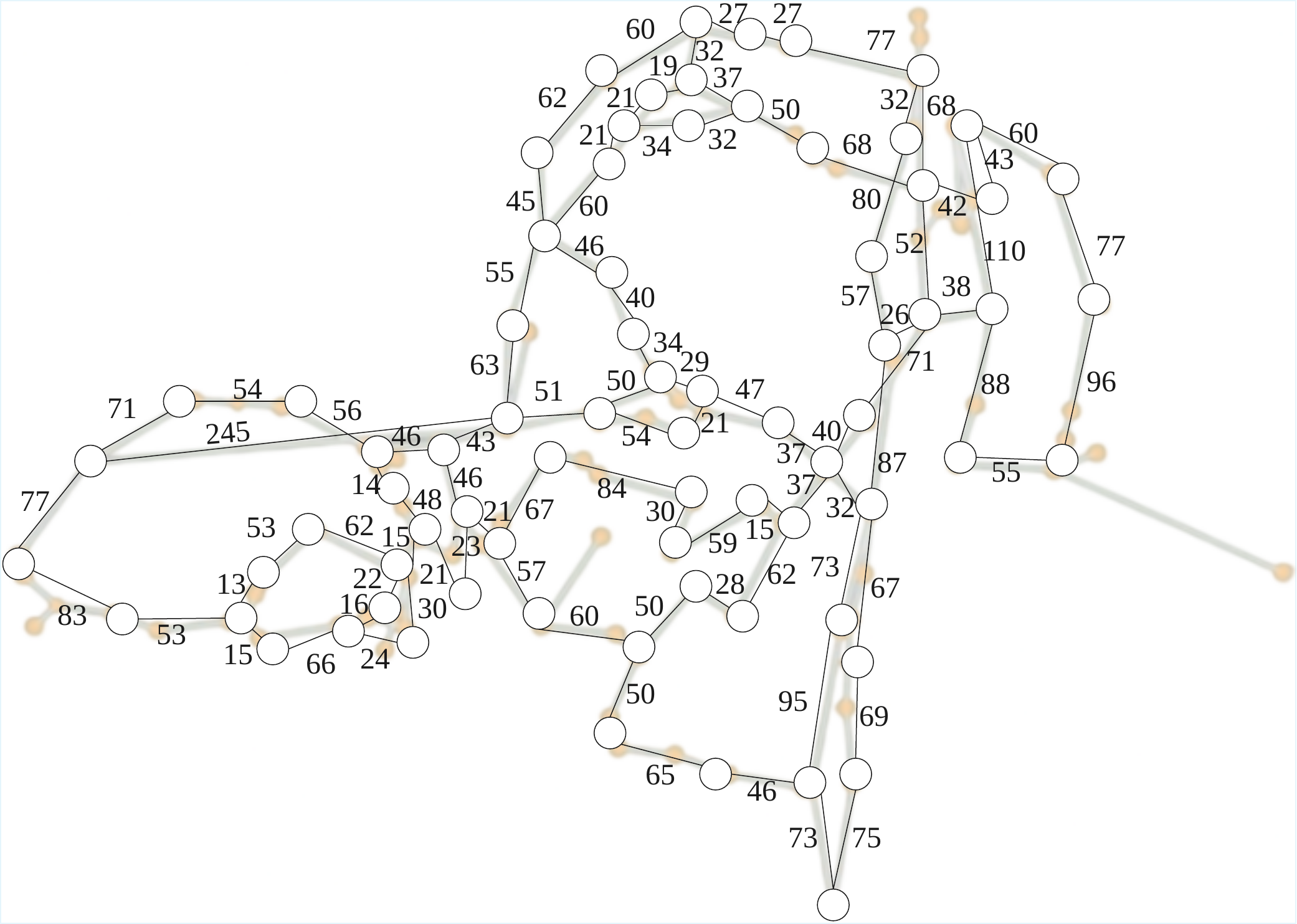}
        \label{fig:topologyION}
    }
    \subfloat[Catalunya topology \cite{catalunya}.]{
        \includegraphics[width=0.38\textwidth]{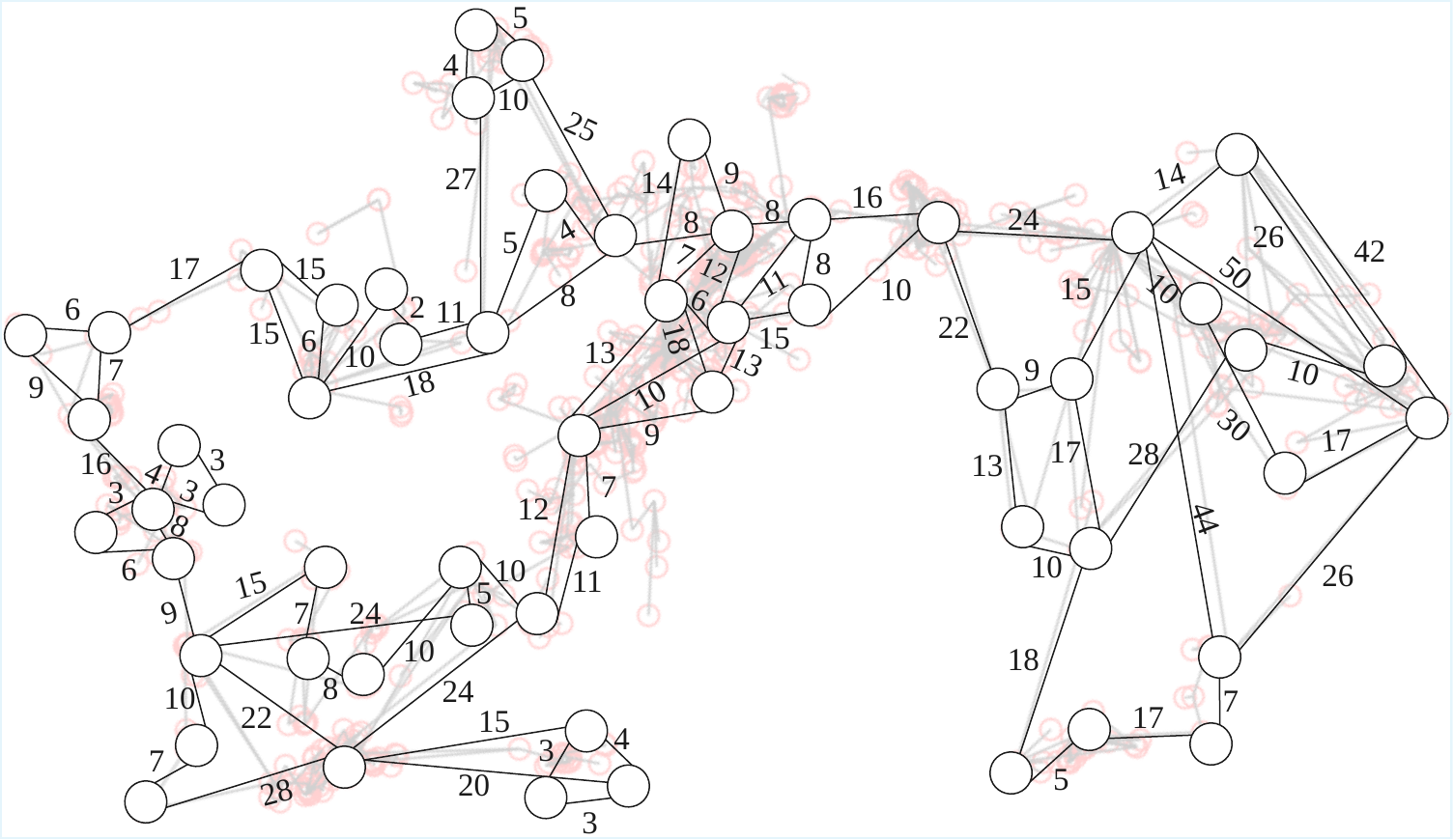}
        \label{fig:topologyCatalunya}
    }
        \subfloat[Data center topology \cite{ranjith2023performance}.]{
        \includegraphics[width=0.25\textwidth]{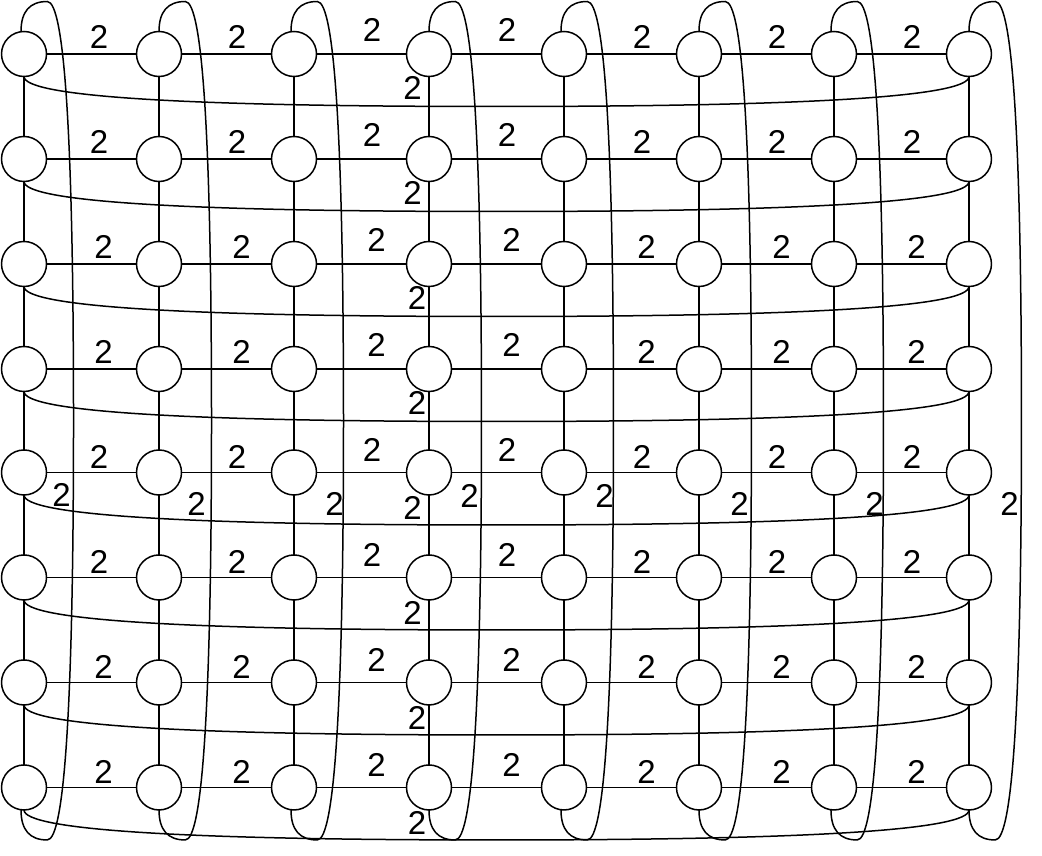}
        \label{fig:topologyDC}
    }
    \caption{The studied topologies with link lengths in km.} \label{fig: topos}
\end{figure*}

Given a network represented by a graph $G=(V,E)$, where $|V|$ and $|E|$ denote the number of nodes and links in the graph, respectively, the GSPC algorithm requires the set $\mathcal{P}$ containing the shortest path for each source-destination node pairs, and a maximum of $N$ DFOS devices to be deployed in the network. The objective is to select a subset of links so that each shortest end-to-end path is monitored by at least one DFOS-enabled link, until either all $N$ available DFOS devices are deployed in the network or all links are sensed. The algorithm first assigns a unique identifier $p$ to each shortest path (line 1 in Algorithm \ref{alg:sp_coverage}) and constructs a link-path incidence mapping $\mathcal{M}$ that associates each link with the set of shortest paths traversing it (line 2). All shortest paths are initially marked as unsensed $\mathcal{U}$ (line 3), and an empty list of sensed links $\mathcal{L}$ is created (line 4). At each iteration (line 5), the GSPC evaluates every link in the network and computes its coverage $cover(e)$ (line 9), defined as the number of currently unsensed shortest paths that include that link. The link with the highest coverage $e^\star$ is selected (lines 10-12), and the corresponding value of the link's $e^\star$ coverage is then stored in $bestCover$ (line 11). A DFOS device is assigned to the link, and the link is added to the list $\mathcal{L}$ (line 13). All shortest paths traversing the selected link $e^\star$ are then marked as sensed (line 13) and removed from further consideration (line 14). This process is repeated until either all shortest paths are sensed or the number of DSF devices $N$ is reached (line 5).

\begin{algorithm}[h!]
\caption{Greedy Shortest-Path Coverage Link Selection}
\label{alg:sp_coverage}
\begin{algorithmic}[1]
\Require
Graph $G=(V,E)$; set of all possible shortest paths $\mathcal{P}$; number of DFOS devices $N$
\Ensure
Selected sensing links $\mathcal{L}$

\State Assign a unique identifier to each path $p \in \mathcal{P}$
\State Build a mapping $\mathcal{M}: e \rightarrow \{p \mid p \in \mathcal{P}, e \in p\}$

\State $\mathcal{U} \leftarrow \mathcal{P}$ \Comment{Set of unsensed shortest paths}
\State $\mathcal{L} \leftarrow \emptyset$ \Comment{Selected sensing links}

\While{$\mathcal{U} \neq \emptyset$ $\And$ $|\mathcal{L}| < N$}
    \State $bestCover \leftarrow 0$
    \State $e^\star \leftarrow \text{null}$

    \ForAll{$e \in E$}
        \State $cover(e) \leftarrow |\mathcal{M}(e) \cap \mathcal{U}|$
        \If{$cover(e) > bestCover$}
            \State $bestCover \leftarrow cover(e)$
            \State $e^\star \leftarrow e$
        \EndIf
    \EndFor

    \State $\mathcal{L} \leftarrow \mathcal{L} \cup \{e^\star\}$
    \State $\mathcal{U} \leftarrow \mathcal{U} \setminus \mathcal{M}(e^\star)$
\EndWhile
\Return $\mathcal{L}$

\end{algorithmic}
\end{algorithm}
The computational complexity of Algorithm~\ref{alg:sp_coverage} is governed by the iterative link evaluation process. Let $ L = \sum_{p \in \mathcal{P}} |p| $ denote the total number of path-link incidences across all shortest paths. The preprocessing phase, in which the KSP list $\mathcal{P}$ and the link-path incidence mapping $\mathcal{M}$ (line 2) are constructed, requires $O(V^2(E + V \log V))$ and  $O(L)$ time, respectively.
The subsequent selection phase consists of a greedy loop that executes for at most $N$ iterations, where $N$ denotes the maximum number of DFOS-enabled links to be deployed. During each iteration, the algorithm evaluates all links and recomputes their coverage with respect to the set of unsensed shortest paths. Since the total size of the incidence mapping is $L$, each iteration requires $O(L)$ time. Consequently, the overall complexity of the selection phase is $O(NL)$. Therefore, the total time complexity of the algorithm is $O(V^2(E + V log V) + L + NL) = O(V^2(E + V log V) + NL)$. In the worst case, each shortest path may traverse up to $|V|$ nodes, yielding $L = O(|V|\,|\mathcal{P}|)$. The resulting worst-case complexity is therefore $O(V^2(E + V log V) + N\,|V|\,|\mathcal{P}|)$. Since the algorithm scales polynomially with the number of links and shortest paths, it remains computationally tractable for large-scale network topologies and is significantly more efficient than exhaustive search approaches.

\section{performance evaluation}\label{sec:perfEval}

\subsection{Simulation Setup} 

We used the optical network simulator~\footnote{https://ons-simulator.com/} to model four event types: service requests, failure prediction, failure occurrence, and recovery. All results are presented with 95\% confidence intervals., and each load point is plotted as the average result of 10 simulations with different seeds. Each simulation has 100,000 circuit requests that arrive according to a Poisson process and are uniformly distributed across all node pairs. Request holding times follow a negative exponential distribution with an average duration of 600 s, while bandwidth demands are uniformly generated between 10 and 400 Gbps in 10 Gbps increments. Two modulation formats are supported, namely 8 QAM and 16 QAM, providing sub-carrier capacities of 37.5 Gbps and 50 Gbps, respectively, with maximum transmission reaches of 1000 km and 500 km \cite{shakouri2021proactive}. The modulation format selected for each request is the most spectrally efficient which satisfies the corresponding path length. Each link is equipped with 320 spectral slots, and a one-slot guard band (12.5 GHz) is reserved between adjacent optical circuits. Routing is performed using the $k$-shortest paths algorithm with $k=3$ for all source–destination pairs, and spectrum allocation follows a \textit{first-fit} policy.

The overall simulation time is assumed to correspond to one year of network operation. To model failures, 100 link failures are generated uniformly over time, each with a mean time to repair (MTTR) of 4 hours. Failure prediction is assumed to occur with a 6-minute anticipation window \cite{mazur2024real}. Additionally, we analyzed the impact of different DFOS devices with varying sensing responsiveness on network performance. After a link failure, circuit reallocation attempts are constrained to a maximum duration of 20 ms \cite{sambo2020locally}. Sensing is limited to links up to 100 km in length, as the considered metropolitan and data center networks do not include optical amplifiers. The ST and SWT metrics are computed according to Eqs. (\ref{eq:setuptime}), and (\ref{eq:pst_restoration}), assuming D = 10~$\mu$s, P = 400~$\mu$s, and F = 500~$\mu$s \cite{765615} when sensing is not employed. In sensing-enabled scenarios, an additional response time of 340~ns from the sensing device is added to F \cite{szczerban2024optical}.

We simulate two metro and one data center topologies, as illustrated in Fig.~\ref{fig: topos}: the ION topology \cite{ion}, presenting a multi-ring metro network from New York City with 74 nodes of $2.89$ average degree, and 184 unidirectional links with average length of $40.71$ km; and a simplified version of the Catalunya topology \cite{catalunya}, presenting a smaller and more connected multi-ring metro network from Catalunya City with 57 nodes, $2.96$ average node degree, and 172 unidirectional links of average $13.27$ km. The chosen data center network is a NovaCube (2D-torus) topology, which represents the interconnection in super-computing clusters and modern AI acceleration pods \cite{ranjith2023performance},  with 64 nodes, 4 average node degree, and 256 unidirectional links with average distance of 2~km \cite{xie2022scaling}. 
For the evaluation, we compare the previously presented heuristics by using the following performance measures: (i) Bandwidth blocking rate (BBR), represents the percentage of blocked circuits out of the total number of circuits requested; (ii) Affected circuits, represents the number of circuits that get disrupted by a failure event, regardless of whether they were successfully recovered after a failure or not; (iii) Prediction time (ms), to represent the time between a failure event prediction and its occurrence; and (iv) Downtime, overall unavailability of all circuits affected by a failure event.

\subsection{Results and Discussion} 
\begin{figure*}[b]
    \centering
    \subfloat[\footnotesize BBR in different DFOS placement heuristics in ION. ]{
        \includegraphics[width=0.35\textwidth]{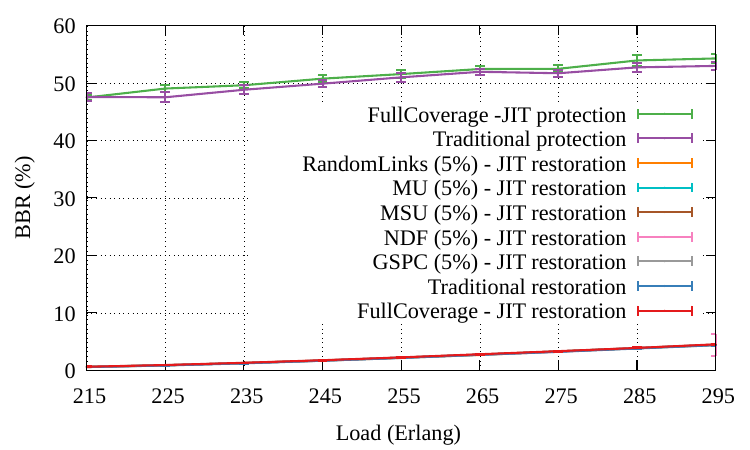}
        \label{fig:IONBBR}
    }
    \subfloat[ \footnotesize BBR in different DFOS placement heuristics in data center. ]{
        \includegraphics[width=0.35\textwidth]{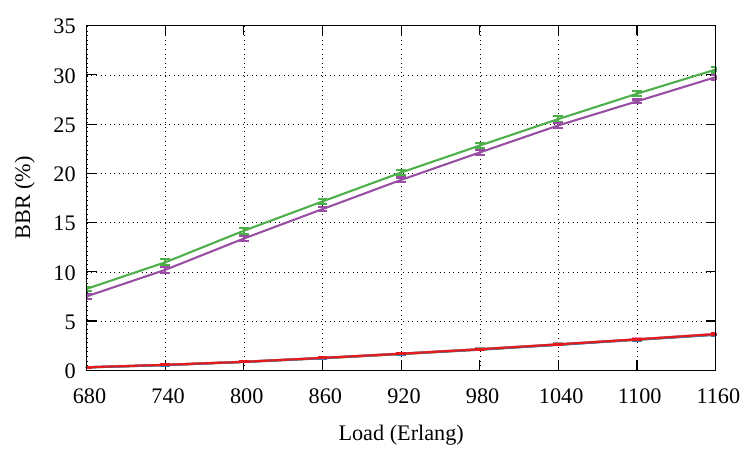}
        \label{fig:DCBBR}
    }
    \centering
    \caption{Impact of failures on BBR in both ION and data center topologies. (5\%): 5\% of the total links are sensed (monitored).} \label{fig:BBRBOTH}
\end{figure*}

\begin{figure*}[h]
    \centering
    \subfloat[\footnotesize Impact of varying the number of DFOS devices \\with different heuristics on the affected circuits.]{
        \includegraphics[width=0.33\textwidth]{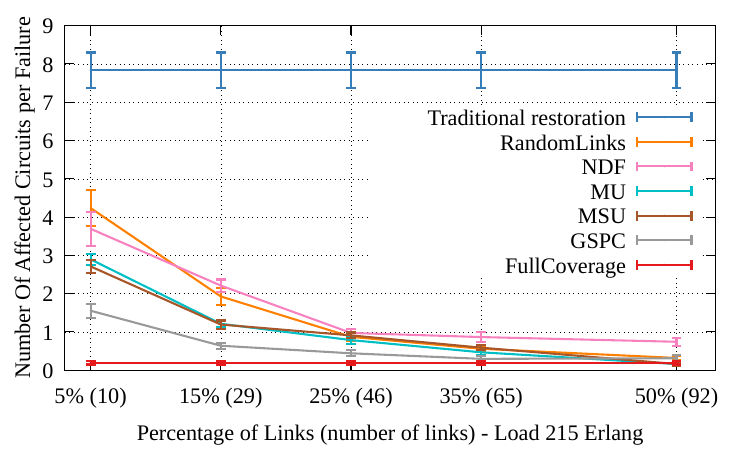}
        \label{fig:IONDFOS_percentage}
    }
    \subfloat[\footnotesize Impact of varying the prediction time with\\ different heuristics on the affected circuits.]{
        \includegraphics[width=0.33\textwidth]{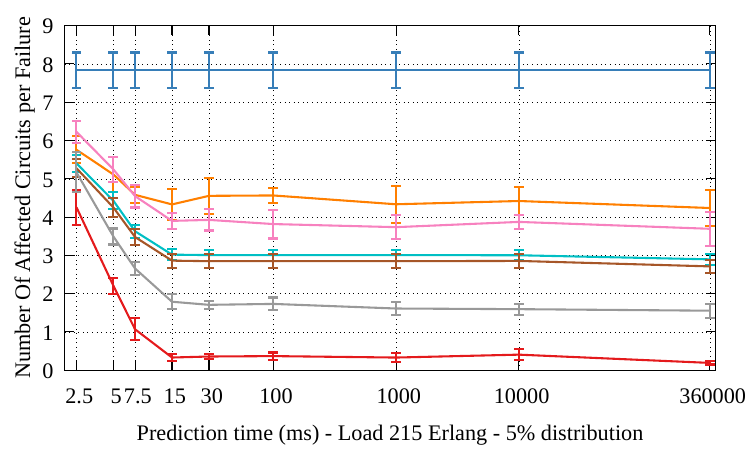}
        \label{fig:IONsensing_vs_heuristics}
    }
 \subfloat[\footnotesize Impact of varying the prediction time with\\ different heuristics on the downtime.]{        \includegraphics[width=0.33\textwidth]{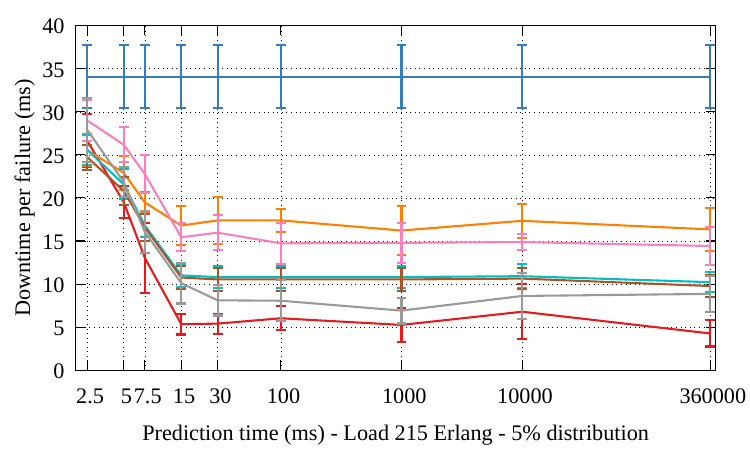}
        \label{fig:IONdowntime}
    }
    \caption{Impact of failures on the affected circuits and downtime with different DFOS placement heuristics in ION.}    \label{fig:ion_results}\label{fig:ION-metrics}
\end{figure*}

\begin{figure*}[h]
    \centering
    \subfloat[\footnotesize Impact of varying the number of DFOS devices\\ with different heuristics on the affected circuits.]{
        \includegraphics[width=0.33\textwidth]{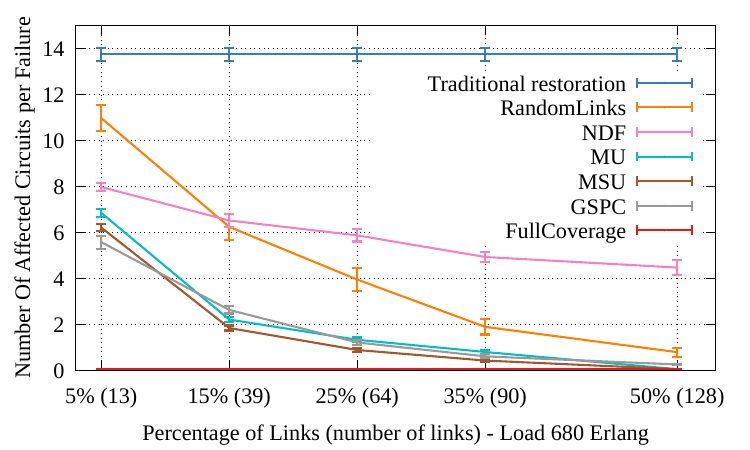}
        \label{fig:dc_affected_lps}
    }
    \subfloat[\footnotesize Impact of varying the prediction time with\\ different heuristics on the affected circuits.]{
        \includegraphics[width=0.33\textwidth]{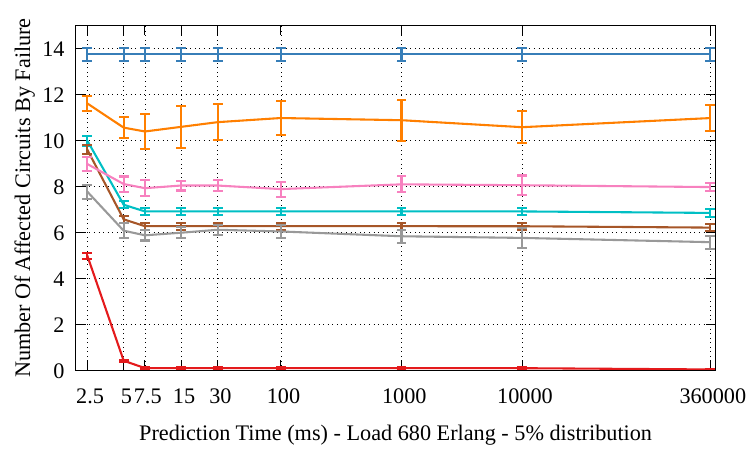}
        \label{fig:dc_DFOS_affected}
    }
 \subfloat[\footnotesize Impact of varying the prediction time with\\ different heuristics on the downtime.]{        \includegraphics[width=0.33\textwidth]{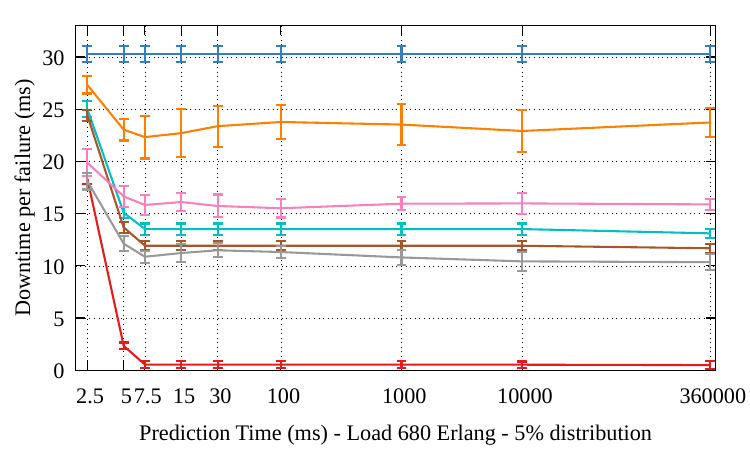}
        \label{fig:DC_Downtime}
    }
    \caption{Impact of failures on the affected circuits and downtime with different DFOS placement heuristics in data center.}
    \label{fig:dc_results}\label{fig:DC-metrics}
\end{figure*}

We begin by analyzing the BBR shown in Fig.~\ref{fig:BBRBOTH}. The chosen load values are selected such that the restoration BBR remains below 5\%. As expected, traditional and full coverage protection are highly inefficient (an average of 50\% for ION, 20\% for DC) due to the dedicated allocation of idle backup resources, whereas all restoration heuristics achieve blocking levels below 5\%. Furthermore, despite reserving one fixed spectrum slot per link in all heuristics where DFOS is applied, the associated overhead is negligible and does not impact the BBR.

Now, utilizing a load that leads to a BBR $< 1\%$ (around 0.8\%), we investigate how efficient different heuristics are when choosing the placement of DFOS devices, in Fig.~\ref{fig:ION-metrics}(a) for ION and in Fig.~\ref{fig:DC-metrics}(a) for the DC, leveraging the 6 minutes suggested in \cite{mazur2024real} for the prediction time window. We take traditional restoration (without DFOS, thus reactive) as the upper bound and a full-coverage scenario (where all links can be sensed) as the lower bound.  With only 5\% unidirectional links (10  for ION and 13 for DC) being sensed, we achieve a reduction of $45.9\%$ (ION) and $20.1\%$ DC with the random links heuristic when compared to the upper bound. However, when we assess the impact of the GSPC, with only 5\% of the links utilizing DFOSs, it achieves an immense reduction of $81.3\%$ (ION) and $59.4\%$ (DC), thus closing on the gap of a full coverage scenario. As the number of links with DFOS increases, the number of affected circuits declines, but stabilizes at roughly 25\% coverage (46 links) for ION and (64 links) for DC. At this point,  the GSPC heuristic achieves a $95.1$\% reduction (with 0.37 affected circuits per failure) for ION and a  $91.2\%$ reduction (1.47 affected circuits) for DC, compared to a full coverage scenario with affected circuits per failure of 0.19 and 0.11 for ION and DC, respectively. 

Furthermore, Figs. \ref{fig:ION-metrics} and \ref{fig:DC-metrics} illustrate the impact of deploying DFOS devices on 5\% of the unidirectional network links with different prediction times, in terms of the number of affected circuits (b) and the total service  downtime (c), with 360000~ms (6~mins) being the upper-limit provided in \cite{mazur2024real}. Both sub-figures \ref{fig:ION-metrics}(b) and \ref{fig:DC-metrics}(b) present the average number of affected circuits as a function of failure event prediction time provided by different DFOS devices in ION and DC, respectively. The lowest curve (red) presents a full coverage scenario, and the highest (blue) presents the traditional restoration. Results indicate that a short prediction time ($\leq 15$~ms) significantly reduces the number of affected circuits. For instance, in ION, enabling 5~ms of prediction time can, on one hand, reduce the number of affected circuits, when compared to traditional restoration, by $55.4\%$ for GSPC and by $34.7\%$ for random links. On the other hand, compared with full network sensing coverage, GSPC shows $55.8\%$ more affected circuits, whereas random placement results in a $131.7\%$ increase. Similarly, for DC, enabling a 2.5~ms prediction window can decrease the number of affected circuits, when compared to traditional restoration, by $43.4\%$ for GSPC and by $15.5\%$ for random links. Also, the increase in affected circuits is limited by $55.9\%$ for GSPC (best-heuristic), whereas the random links heuristic results an increment of  $133.1\%$ compared to the full coverage sensed network, thus demonstrating the superior performance of the GSPC heuristic in limiting the number of affected circuits. As prediction time increases, the number of affected circuits declines, but stabilizes after roughly 15~ms (ION) and 7.5~ms (DC). At these points, GSPC in ION achieves 1.78 affected circuits per failure (compared to 0.23 for the full coverage baseline) and  $5.9$ affected circuits in DC ($0.09$ for the full coverage baseline). Beyond this point, additional prediction time yields negligible improvements. 

Fig.~\ref{fig:ION-metrics}(c) and \ref{fig:DC-metrics}(c) show the average downtime per failure event for ION and DC, respectively, when sensing is enabled on only 5\% of the links. Increasing the prediction time consistently reduces overall downtime. For example, even with the random links heuristic, a short sensing interval of 2.5 ms reduces average downtime by $8.5~ms$ ($25.1\%$) in ION and $3.1~ms$ ($9.8\%$) in DC compared to traditional restoration. As the prediction time increases, downtime continues to decrease in both topologies. At 15 ms, random links achieve a downtime reduction of $17.2~ms$ ($50.7\%$) in ION, while GSPC reaches $20.3~ms$ ($78.6\%$) reduction. For DC at 5 ms, random links reduce downtime by $7.2~ms$ ($23.8\%$), and GSPC achieves an $18.2~ms$ ($59.8\%$) reduction. When compared to the full coverage scenario, the increase in service downtime for GSPC is limited by $89.3\%$ in ION at 15 ms and by $419.3\%$ in DC at 5 ms. Beyond these prediction times for each topology, additional prediction time yields negligible benefits.

\begin{table}[h!]
\centering
\small
\setlength{\tabcolsep}{5pt}
\renewcommand{\arraystretch}{1.2}
\begin{tabular}{>{\raggedright\arraybackslash}p{3cm}cccc}
\toprule
\textbf{Metric} & \multicolumn{2}{c}{\textbf{ION ~~| Catalunya}} & \multicolumn{2}{c}{\textbf{ION ~~| Catalunya}} \\
\cmidrule(lr){2-3} \cmidrule(lr){4-5}
 & \multicolumn{2}{c}{\hspace{1pt}{JIT Protection}} & \multicolumn{2}{|c}{\hspace{1pt}{JIT Restoration}} \\
\midrule

BBR (\%) & 51.29 & 54.17 & ~2.36 & 3.16 \\

\midrule
\multicolumn{5}{c}{\textbf{Monitored Links}} \\

\multicolumn{1}{l}{} & \multicolumn{2}{c}{\textbf{5\%}} & \multicolumn{2}{|c}{\textbf{25\%~~}} \\
Affected circuits (RL) & 4.23 & 4.78 & ~0.90 & 1.14 \\
Affected circuits (GSPC) & 1.55 & 1.21 & ~0.44 & 0.33 \\

\midrule
\multicolumn{5}{c}{\textbf{Prediction Time (ms) with 5\% Monitored Links}} \\

\multicolumn{1}{l}{} & \multicolumn{2}{c}{\textbf{ 5 ms}} & \multicolumn{2}{|c}{\textbf{15 ms~~~}} \\
Affected circuits (RL) & 5.11 & 5.87 & ~4.33 & 5.36 \\
Affected circuits (GSPC) & 3.49 & 2.80 & ~1.79 & 1.21 \\

\multicolumn{1}{l}{} & \multicolumn{2}{c}{\textbf{ 2.5 ms}} & \multicolumn{2}{|c}{\textbf{ 15 ms~~~~}} \\
Downtime (ms) (RL) & 25.5 & 24.2 & ~16.7 & 17.5 \\
Downtime (ms) (GSPC) & 27.8 & 20.1 & ~10.1 & 6.26 \\

\bottomrule
\end{tabular}
\caption{Performance comparison of protection and restoration for random links (RL) and GSPC DFOS placement heuristics.}
\label{tab:metrics}
\end{table}

Due to the similar performance metrics observed across both metropolitan topologies, results for the Catalunya topology are presented only in Table \ref{tab:metrics}. The table is divided into three parts, in which the BBR (from Fig.~\ref{fig:BBRBOTH}) is compared between the ION and Catalunya topologies in the first part. The second shows the impact of the number of distributed DFOS on the heuristic solutions that perform the worst (random link) and best (GSPC) on the number of affected circuits, as illustrated in Fig.~\ref{fig:ION-metrics}(a). The last part examines the impact of varying the prediction time on affected circuits (Fig.~\ref{fig:ION-metrics}(b)) and on downtime (Fig.\ref{fig:ION-metrics}(c)). The comparison reveals the same behavior in both topologies, highlighting the importance of distributing DFOS across the network and the impact of the placement strategy. Even with DFOS deployed on only 5\% of all links, a random placement can still mitigate the impact of failures. In contrast, the GSPC heuristic provides substantially greater benefits by selecting links that cover a larger number of relevant shortest paths.


\section{Conclusion}\label{sec:conc}

In this paper, we proposed an architecture for the distributed fiber optical sensing (DFOS)-enabled system architecture for elastic optical networks (EONs), based on the ITU-T standards. Furthermore, we investigate the impact of different DFOS device placement strategies in metropolitan and data center EONs on various network performance metrics, like bandwidth blocking probability, number of circuits affected by a failure event, and the service downtime.  In particular, we focus on the case with only 5\% of the unidirectional network links equipped with DFOS devices, representing the minimum DFOS deployment considered in our study, and evaluate the prediction time required for DFOS to detect an impending failure and enable Just-In-Time (JIT) restoration before the failure occurs.
Our results illustrate that, when utilizing the Greedy Shortest Path Coverage (GSPC) heuristic, equipping only 5\% of the links with DFOS devices can achieve substantial reductions in the number of affected circuits, while increasing the deployment to 25\% yields results comparable to those obtained with full network coverage. For the 5\% deployment scenario, we further show that prediction windows of 15~ms for metropolitan networks and 7.5~ms for data center networks are sufficient to enable JIT restoration and mitigate failures impact. 
Our findings open new directions for future research, including the investigation of multi-failure scenarios, false alarms, missed detections, and how to efficiently associate DFOS placement heuristics with distinct routing solutions.



\bibliography{mybib}

@inproceedings{ranjith2023performance,
  title={Performance Study of Three Different Types of Data Center Network Architectures, NovaCube, BCube, and FatTree Using NS-3 Simulator},
  author={Ranjith, A and Subhash, P and Nagaraju, G and Srinivas, M and Prashanth, A},
  booktitle={International Conference on Computer \& Communication Technologies},
  pages={411--421},
  year={2023},
  organization={Springer}
}

@article{xie2022scaling,
  title={Scaling optical interconnects for hyperscale data center networks},
  author={Xie, Chongjin and Zhang, Bo},
  journal={Proceedings of the IEEE},
  volume={110},
  number={11},
  pages={1699--1713},
  year={2022},
  publisher={IEEE}
}

@INPROCEEDINGS{10901306,
  author={Zhu, Julie Yixuan and et. al.},
  booktitle={GLOBECOM 2024 - 2024 IEEE Global Commun. Conf.}, 
  title={Semantic FaultAware (SFA): Proactive Fault Prediction Architecture in SD-WAN with Semantic Knowledge Graphs and Machine Learning}, 
  year={2024},
  volume={},
  number={},
  pages={5096-5101},
  doi={10.1109/GLOBECOM52923.2024.10901306}}

@INPROCEEDINGS{765615,
  author={Ramamurthy, S. and Mukherjee, B.},
  booktitle={1999 IEEE Int. Conf. on Commun. (Cat. No. 99CH36311)}, 
  title={Survivable WDM mesh networks. II. Restoration}, 
  year={1999},
  volume={3},
  number={},
  pages={2023-2030 vol.3},
  doi={10.1109/ICC.1999.765615}}

@article{shakouri2021proactive,
  title={Proactive restoration with low re-provisioning in Elastic Optical Networks},
  author={Shakouri, Saba and et. al.},
  journal={Comput. Netw.},
  volume={197},
  pages={108284},
  year={2021},
  publisher={Elsevier}
}

@inproceedings{mazur2024real,
  title={Real-Time Monitoring of Cable Break in a Live Network using a Coherent Transceiver Prototype},
  author={Mazur, Mikael and et. al.},
  booktitle={2024 Optical Fiber Commun. Conf. and Exhibition (OFC)},
  pages={1--3},
  year={2024},
  organization={IEEE}
}

@inproceedings{szczerban2024optical,
  title={Optical Fiber Sensing Network Control Plane Enabled by a Novel Sub $\mu$s Response Time Fiber Sensing Control Device},
  author={Szczerban, Mijail and et. al.},
  booktitle={Opt. Fiber Commun. Conf.},
  pages={},
  year={2024},
  organization={Optica Publishing Group}
}

@INPROCEEDINGS{catalunya,
  author={Vega, Davide and et. al.},
  booktitle={2012 IEEE 8th Int. Conf. on Wireless and Mobile Comput., Netwo. and Commun. (WiMob)}, 
  title={Topology patterns of a community network: Guifi.net}, 
  year={2012},
  volume={},
  number={},
  pages={612-619},
  doi={10.1109/WiMOB.2012.6379139 }}

@misc{ion,
	author = {topology-zoo},
	title = {ion},
	howpublished = {\url{https://web.archive.org/web/20220308153457/https://topology-zoo.org/maps/Ion.jpg}},
	year = {},
	note = {[Accessed 02-11-2025]},
}

@article{sambo2020locally,
  title={Locally automated restoration in SDN disaggregated networks},
  author={Sambo, Nicola},
  journal={IEEE J. Opt. Commun. Netw.},
  volume={12},
  number={6},
  pages={C23--C30},
  year={2020},
  publisher={IEEE}
}

@article{wellbrock2023explore,
  title={Explore benefits of distributed fiber optic sensing for optical network service providers},
  author={Wellbrock, Glenn A and et. al.},
  journal={Journal of Lightwave Technology},
  volume={41},
  number={12},
  pages={3758--3766},
  year={2023},
  publisher={IEEE}
}

@article{he202550,
  title={50 ms optical-layer failure recovery via fast-switching wavelength-selective switches and segment-based shared protection},
  author={He, Junyue and et. al.},
  journal={IEEE J. Opt. Commun. Netw.},
  volume={17},
  number={11},
  pages={967--983},
  year={2025},
  publisher={IEEE}
}

@article{ye2022distributed,
  title={Distributed fiber optic sensors placement for infrastructure-as-a-sensor},
  author={Ye, Zilong and Ji, Philip N and Wang, Ting},
  journal={Photonic Network Communications},
  volume={44},
  number={2},
  pages={90--101},
  year={2022},
  publisher={Springer}
}

@inproceedings{ye2021survivable,
  title={Survivable distributed fiber optic sensors placement against single link failure},
  author={Ye, Zilong and et al.},
  booktitle={ICC 2021-IEEE International Conference on Communications},
  pages={1--6},
  year={2021},
  organization={IEEE}
}

@article{zhang2026integrated,
  title={Integrated sensing and communications over the years: An evolution perspective},
  author={Zhang, Di and Cui, Yuanhao and Cao, Xiaowen and Su, Nanchi and Gong, Yi and Liu, Fan and Yuan, Weijie and Jing, Xiaojun and Zhang, J Andrew and Xu, Jie and others},
  journal={IEEE Communications Surveys \& Tutorials},
  year={2026},
  publisher={IEEE}
}

@INPROCEEDINGS{mouammar:26,
  author={Mouammar, Sleman and Brasileiro, \'Italo B. and Drummond, André C.},
  booktitle={ICC 2026 - IEEE International Conference on Communications}, 
  title={Just-in-time Restoration with Distributed Fiber Sensing in Metropolitan Optical Networks}, 
  year={2026},
  volume={},
  number={},
  pages={1-6},
  doi={10.1109/ICC59461.2026.11587436}}

@ARTICLE{11424028,
  author={He, Haijun and Li, Peng and Fan, Hailong and Pu, Zhengyu and Zou, Xihua and Pan, Wei and Yan, Lianshan},
  journal={Journal of Optical Communications and Networking}, 
  title={Fiber-based photonic integrated sensing and communication: enabling technologies and key challenges}, 
  year={2026},
  volume={18},
  number={4},
  pages={B33-B48},
  doi={10.1364/JOCN.584969}}

@techreport{itu2025g681,
author = {{International Telecommunication Union}},
title = {{G.681: Distributed fibre optic sensing system for terrestrial optical transmission systems}},
institution = {International Telecommunication Union},
year = {2025},
month = nov,
number = {ITU-T Recommendation G.681},
type = {Recommendation},
url = {https://www.itu.int/rec/T-REC-G.681-202511-I}
}

@ARTICLE{11435207,
author={Kotrla, Scott R. and Huang, Ming-Fang and Fang, Jian and Han, Shaobo and Lynn, Jamie and Ip, Ezra and Mundt, Jeffrey A. and Wang, Ting},
journal={Journal of Optical Communications and Networking}, 
title={Leveraging deployed telecom cables for distributed fiber sensing topologies and applications}, 
year={2026},
volume={18},
number={4},
pages={B72-B84},
doi={10.1364/JOCN.585523}
}

@article{ambrosone2026enabling,
title={Enabling real-time optical networks as distributed sensors via open-source telemetry architecture},
author={Ambrosone, Renato and Srivallapanondh, Sasipim and Schips, Riccardo and Malik, Gulmina and Straullu, Stefano and Aquilino, Francesco and Nespola, Antonino and Virgillito, Emanuele and Napoli, Antonio and Clivati, Cecilia and others},
journal={Journal of Optical Communications and Networking},
volume={18},
number={6},
pages={B187--B199},
year={2026},
publisher={Optica Publishing Group}
}

@article{chaves2012novel,
title={Novel strategies for sparse regenerator placement in translucent optical networks},
author={Chaves, Daniel AR and Carvalho, Renan VB and Pereira, Helder A and Bastos-Filho, Carmelo JA and Martins-Filho, Joaquim F},
journal={Photonic Network Communications},
volume={24},
number={3},
pages={237--251},
year={2012},
publisher={Springer}
}

@article{wang2026sensing,
  title={Sensing as a service: can in-band distributed fiber optic sensing coexist with coherent metro DWDM communication network?},
  author={Wang, Jingchuan and Lau, Alan Pak Tao and Lu, Chao},
  journal={Journal of Lightwave Technology},
  volume={44},
  number={5},
  pages={1712--1720},
  year={2026},
  publisher={IEEE}
}


\end{document}